\documentclass[aps,pre,preprint,superscriptaddress,showpacs]{revtex4}
\usepackage{amssymb}
\usepackage{epsfig}
\usepackage{amsmath}
\usepackage{times}
\usepackage{color}
\usepackage{lipsum}
\usepackage{lineno}
\begin{document}
\title{Quorum sensing in populations of spatially extended chaotic oscillators \\ coupled indirectly via a heterogeneous environment}
\author{Bing-Wei Li}
\email[To whom correspondence should be addressed. Email address: ]{bwli@hznu.edu.cn}
\affiliation{Department of Physics, Hangzhou Normal University,
Hangzhou 310036, China}
\author{Xiao-Zhi Cao} \affiliation{Department of Physics, Hangzhou Normal University,
Hangzhou 310036, China}
\author{Chenbo Fu} \affiliation{Department of Automation, Zhejiang University of Technology, Hangzhou 310023, China}

\begin{abstract} 
Many biological and chemical systems could be modeled by a population of oscillators coupled indirectly via a dynamical environment. Essentially, the environment by which the individual elements communicate is \emph{heterogeneous}. Nevertheless, most of previous works considered the homogeneous case only. Here, we investigated the dynamical behaviors in a population of spatially distributed chaotic oscillators immersed in a heterogeneous environment. Various dynamical synchronization states such as oscillation death, phase synchronization, and complete synchronized oscillation as well as their transitions were found. More importantly, we uncovered a non-traditional quorum sensing transition: increasing the density would first lead to collective oscillation from oscillation quench, but further increasing the population density would lead to degeneration from complete synchronization to phase synchronization or even from phase synchronization to desynchronization. The underlying mechanism of this finding was attributed to the dual roles played by the population density. Further more, by treating the indirectly coupled systems effectively to the system with directly local coupling, we applied the master stability function approach to predict the occurrence of the complete synchronized oscillation, which were in agreement with the direct numerical simulations of the full system. The possible candidates of the experimental realization on our model was also discussed.
\end{abstract}
\date{\today }
 \maketitle

\section{Introduction}
Synchronization, firstly discovered by Huygens at the least 300 years ago, has been recognized as a universal concept in the realm of the nonlinear science \cite{pikovsky:book,Strogatz:book}. The synchronized motion is of fundamental importance in coordinating the rhythmic behavior among individuals in various systems ranging from physics, chemistry to biology \cite{kuramoto,glass,goldbeter,winfree}. Well-known examples include the arrays of lasers \cite{vladimirov}, Jsoephson junction series \cite{wiesenfeld}, assembles of chemical oscillators \cite{toiya}, cardiac muscle cells \cite{bers_nat} and neurons in brain \cite{belykh}. In the context of cardiovascular science, synchronous contraction of the heart is essential to pump blood throughout the whole body, while asynchronous contraction of the heart may lead to serious cardiac arrhythmias \cite{panfilov_sci,gray_nat,witkowski_nat}. In neuroscience, synchronization is believed to be a central
mechanism for neuronal information processing within a brain area and also for communication between different areas of the brain \cite{singer}. On the other hand, the synchronized oscillation may also lead several neurological diseases such as epileptic seizures \cite{mormann} and Parkinson's disease \cite{bergman}.

To investigate the synchronization behaviors in complex systems, a popular as well as an efficient approach is to model the systems as an ensemble of oscillators that are coupled in a direct manner \cite{winfree,kuramoto}. However, in many systems such as bacteria \cite{camilli} , yeast cells \cite{aldridge} and social amoebae {\it Dictyostelium discoideum} \cite{gregor}, the synchronized oscillation is believed to arise through communication by chemical signaling molecules via the extracellular solution. The elements in these systems are not influenced by each other in a direct fashion, but rather indirectly through a common environment. In such kinds of systems, a common finding is that the density of population plays a vital role in determining the dynamical state of the system \cite{monte,ojalvo,Taylorsci,toth,munt,gregor,schwaba_phd12}. For instance, a typical scenario is that as the population density of the element increases and exceeds some threshold value, the system will be suddenly switched from the quiescent state to the state of synchronized oscillation for all the elements. The transition is typically referred to as dynamical ``quorum sensing" (QS). Such dynamical QS transition has also been reported in nonliving systems like a large population of indirectly coupled chemical oscillators \cite{Taylorsci,toth} and lasers \cite{munt}.

Originally, QS was interpreted simply as a means for bacteria to coordinate the collective cellular behaviors within physically and chemically homogeneous cultures. Therefore, in traditional, QS research focused on the well-stirred systems, i.e., they assumed that the concentration of the signaling molecules was distributed uniformly in the external environment \cite{monte,ojalvo,Taylorsci,toth,munt,gregor,schwaba_phd12}. In other words, each element of the system fell the same dynamical environment. But it is now recognized that QS essentially occurs in a complex environment that may be physically, chemically and biologically heterogeneous and under such a condition signaling molecules are transported primarily by the local diffusion \cite{danino,dilanji_jacs12,schutze_bj11,noorbakhsh_pre15,sakaguchi,gouphd,gouphd_jns}. The interaction between the reaction and local diffusion can lead to the emergence of the more complex spatiotemporal patterns compared to the case with homogeneous environment \cite{danino,dilanji_jacs12,schutze_bj11,noorbakhsh_pre15,sakaguchi}.

With systematics investigations, various dynamical synchronization states (e.g., oscillation death (OD), phase synchronization (PS), and complete synchronization oscillation (CSO)) as well as their transitions was uncovered. What is more, a non-traditional quorum sensing transition was observed: increasing the density would first lead to collective oscillation from oscillation quench, but further increasing the parameter of the population density would lead to decrease of degree of synchronization. Specifically, for the small size system, the degeneration of CSO to PS was observed and for large population, transition from  PS to desynchronization occurred. We attributed these new findings to the dual roles of the population density. By treating the indirectly coupled systems effectively to the system with directly local coupling, we applied the master stability function approach to predict the occurrence of the CSO which were in agreement with the direct numerical simulations of the full system.

The remain paper is organized below. In the section II, we made a detailed description of the mathematical model that represents populations of oscillators coupled indirectly through a heterogenous environment. In section III, we studied various synchronous states and their transitions in a small population of chaotic oscillators as function of the population density and diffusion constant. The qualitative as well as a quantitative explanation of the synchronization transition by the Master stability function were  given in section IV. Synchronization and traveling waves were discussed in large population size of the coupled oscillators in section V. We discussed our results and draw a conclusion in section VI and VII, respectively.

\section{The Model and Numerical Method}

\begin{figure}[tbp]
\includegraphics[bb=87pt 374pt 542pt 433pt,clip,scale=1.0]{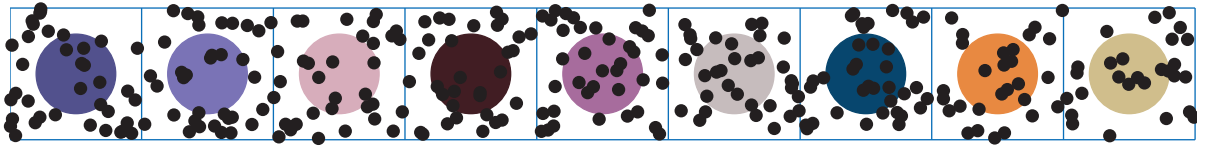} %
\caption{The schematic plot for a group of oscillators indirectly coupled via a diffusive environment. The large filled color dots denote the (R\"{o}ssler) oscillator, and each of them is fixed in the compartment arranged on the line. The small dots represent the signal molecules which are assumed to diffuse freely through the compartments. In addition to the diffusion, these molecules also have interactions with the oscillators.}
\end{figure}

\subsection{A general model}
In the study of QS for a well-stirred system, the position of the individual element in space is of no importance since each element is indirectly coupled in a global way via a homogeneous dynamical environment. However, for the non-stirred system, all the elements are assumed to be fixed in the space, and under such a case the local environment that the element feels may differ from each other; that is to say, the dynamical environment is heterogeneous. Mathematically, such a system could be modeled by a large population of oscillators indirectly coupled via a diffusive environment. A discrete version of this model generally reads \cite{danino,schutze_bj11}
\begin{eqnarray}
\partial_{t}\mathbf{Z}_{i} &=& \textbf{F}(\mathbf{Z}_{i})+K(\mathbf{Z}_{i}^{e}-\mathbf{Z}_{i}),\label{gene_eq1} \\ \partial_{t}\mathbf{Z}_{i}^{e} &=& K\rho(\mathbf{Z}_{i}-\mathbf{Z}_{i}^{e})-J\mathbf{Z}_{i}^{e} + D_{e}\sum_{\left<i,j\right>}(\mathbf{Z}_{j}^{e}-\mathbf{Z}_{i}^{e}).\label{gene_eq2}
\end{eqnarray}
Here, the vector $\textbf{Z}_{i}$ denotes the state (i.e., concentration of intracellular chemical species ) of the element (called oscillator in this paper) that is fixed at the $i$th position in space with $i=1,2,\cdots,N$, and $N$ is the total number of the oscillators. $\mathbf{Z}_{i}^{e}$ represents the extracellular concentration of signaling molecules which are utilized to cell-cell communication. Through our paper, we mean $\mathbf{Z}_{i}^{e}$ by the external environment which the $i$th oscillator feels. Note that due to the heterogeneity of the environment, we here explicitly write the index $i$ for the environment variable $\mathbf{Z}^{e}$.

The parameter $K$ is the coupling strength between the oscillators and the dynamical environment,  and it is assumed to be uniform for all the oscillators. With vanished $K$, the full system is decoupled to two subsystems. One is simply governed by $\partial_{t}\textbf{Z}_{i}=\textbf{F}(\textbf{Z}_{i})$ representing the dynamics of the single oscillator which usually demonstrates limit cycle oscillation. The other subsystem governed by Eq. (\ref{gene_eq2}) represents the dynamical of the environment with the degradation rate $J$. For nonzero $K$, oscillators are indirectly coupled via the environment through $K(\mathbf{Z}_{i}^{e}-\mathbf{Z}_{i})$ and $K\rho(\mathbf{Z}_{i}-\mathbf{Z}_{i}^{e})$. Here $\rho=V_{int}/V_{ext}$, with $V_{int}$ and $V_{ext}$ the intracellular and extracellular volumes, reflects the density of population \cite{li_pre12} and is an important parameter in this paper. With this definition of $\rho$, it is straightforward to see that the exchange of the signals between the oscillators and the environment is balanced. In numerical work, the change of $\rho$ can be achieved by varying the volume of the extracellular with fixed size of the population. The term $D_{e}\sum_{\left<i,j\right>}(\mathbf{Z}_{j}^{e}-\mathbf{Z}_{i}^{e})$ in Eq. (\ref{gene_eq2}) is added to account for the diffusion of signal molecules in the extracellular environment where $D_{e}$ is called diffusion coefficient and $\sum_{\left<i,j\right>}$ means sum of $j$ only with the nearest ones. In the continuum limit, this term is replaced by the Laplacian term \cite{danino}.

The above model, represented by Eqs. (\ref{gene_eq1}-\ref{gene_eq2}), captures the essence of many chemical and biological systems such as the synthetic genetic regulation network used in {\it Escherichia coli} cells \cite{danino}, yeast cell layers \cite{munt} and a dense population of {\it Dictyostelium} cells \cite{noorbakhsh_pre15}. Specifically, for the genetic regulation network, $\partial_{t}\textbf{Z}_{i}=\textbf{F}(\textbf{Z}_{i})$ represents the time evolution of the concentrations of LuxI, AiiA and internal AHL a signaling molecular AHL that can diffuse across the cell membrane and mediates intercellular coupling; while $\textbf{Z}_{e}$ denotes external signaling molecular AHL. It should be noted that the present model assumes an instant coupling between the oscillators and the medium (i.e., no time delay) and that the time scale between the intrinsic dynamics of the oscillators and the external environment are comparable.

The physical picture modeled by Eqs. (\ref{gene_eq1}-\ref{gene_eq2}) for one-dimensional case is illustrated in Fig.~1. In this schematic plot, the large filled color dots denote the oscillators which are fixed with the equal interval in the space. These oscillators are not directly coupled, instead each one only can interact with the local environment, i.e., signaling molecules $\mathbf{Z}^{e}$ which represented by the small black dots. We assume that such small signaling molecules can diffuse freely through the system. It is noted that $\mathbf{Z}_{i}^{e}$ in Eqs. (\ref{gene_eq1}-\ref{gene_eq2}) reflect the average concentration in the $i$th compartment.

\subsection{R\"{o}ssler oscillators coupled via a heterogeneous environment}
As irregular or even chaotic oscillation is ubiquitously observed and it reflects the realistic situations in natural or engineered systems (e.g., the oscillation of the bulk fluorescence is irregular in genetic regulation network used in \cite{danino}), we choose the $\partial_{t}\mathbf{Z}_{i}=\mathbf{F}(\textbf{Z}_{i})$ as chaotic R\"{o}ssler system and use one variable to describe the state of the environmental medium. We further assume that the coupling between the R\"{o}ssler oscillator and environment is via $x$-component. Specifically, the set of equations
we are going to investigate are written,
\begin{eqnarray}
\partial_{t}x_{i}&=&-\omega_{i} y_{i}-z_{i}+K(x^{e}_{i}-x_{i}), \nonumber\\
\partial_{t}y_{i}&=&\omega_{i} x_{i}+ay_{i},\label{rossler}\\
\partial_{t}z_{i}&=&b+(x_{i}-c)z_{i}, \nonumber
\end{eqnarray}
and
\begin{eqnarray}
\partial_{t}x^{e}_{i}=K\rho(x_{i}-x^{e}_{i})-J x^{e}_{i}+\frac{D}{(\Delta x)^{2}}\sum_{\left<i,j\right>}(x_{j}^{e}-x_{i}^{e}) \label{env}.
\end{eqnarray}
In Eqs. (3), the parameter $\omega_i$ represents the intrinsic
frequency of the $i$th oscillator. Here, for the sake of simplicity, we set
$\omega_{i}$ to be identical, $\omega_i$=1. With $(a,b,c)=(0.15,0.4,8.5)$, the isolated oscillator (i.e., $K=0$) shows chaotic oscillation.

Previous works on directly coupled chaotic oscillators have shown that
their collective dynamic was much more complicated and offered even richer phenomena \cite{boccaletti_pr02,pecora_pre90,rosenblum_prl96,rulkov_pre95}. In comparison to directly coupled chaotic oscillators, the works on collective behavior of indirectly coupled oscillators has been much less explored \cite{resmi,singh,chandrasekar}. Particularly, the dynamical QS for chaotic oscillators indirectly coupled through a heterogenous environment remains largely unknown . Our main task in the present work is to study how the collective behavior of the indirectly coupled chaotic oscillators described by Eqs. (\ref{rossler}-\ref{env}) will be varying with the parameters $\rho$ (the population density), and the diffusion strength $D$.

\subsection{Numerical methods and synchronization index}

To numerically integrate Eqs. (\ref{rossler}-\ref{env}), we employ the fourth Rung-Kutta method with a spatial step $\Delta x = \Delta y = 0.2$ and a time step $\Delta t = 0.001$. $J=0.0$ is taken for simplicity. Random initial
conditions are used for the oscillators and a transient period of $t=1.9 \times 10^4$ is discarded when analyzing the properties of the system collective behaviors for the study. The no-flux boundary conditions are employed for the environmental variable $x^{e}$.

To characterize the synchronous transition as functions of parameters, we introduce two synchronization indices to measure the degree of collective oscillation. Following the traditional QS, The first  synchronization index $R$ measuring the degree of the phase synchronization reads \cite{Taylorsci,shinomoto}
\begin{equation}
R =\left<\left|N^{-1}\sum_{j=1}^{N}\exp(i\theta_{j}(t))-\left<
N^{-1}\sum_{j=1}^{N}\exp(i\theta_{j}(t))\right>\right|\right>,
\end{equation}
where $i=\sqrt{-1}$ and $\theta_{j}$ is the phase of the $j$th oscillator defined by $\theta_{j}=tan^{-1}(y_{j}/x_{j})$ in the phase space of $x-y$. $\langle \cdots \rangle$ represents the time average over a
period of $t=1\times 10^3$. It is straightforward to see that if the
oscillators are completely out of phase from each other, we have $R=0$ (Following the tradition, we also set $R=0$ if the oscillators are quenched from oscillation.); while if the oscillators are perfect (phase) synchronization, we have $R=1$.
The second synchronization index is to measure the degree of the complete synchronization (CS), which reads
\begin{equation}
\sigma = \left<\sqrt{\frac{1}{N}\sum_{i=1}^{N}(x_{i}-\overline{x})^2}\right>,~  \overline{x}=\frac{1}{N}\sum_{i}^{N}x_{i}.
\end{equation}
It is straightforward that when CS occurs $\sigma=0$. By involving $R$ and $\sigma$, we can identify various synchronous regimes. For instance, OD suggested by $R=0$ and $\sigma=0$; PS implies $R=1$ but $\sigma$ is a nonzero value; complete synchronous oscillation (CSO) means $R=1$ and $\sigma=0$.

\section{Synchronous states and their transitions on the one dimension chain of a small group of oscillators.}

\begin{figure}[ptb]
\includegraphics[bb=0pt 322pt 401pt 776pt,clip,scale=1.0]{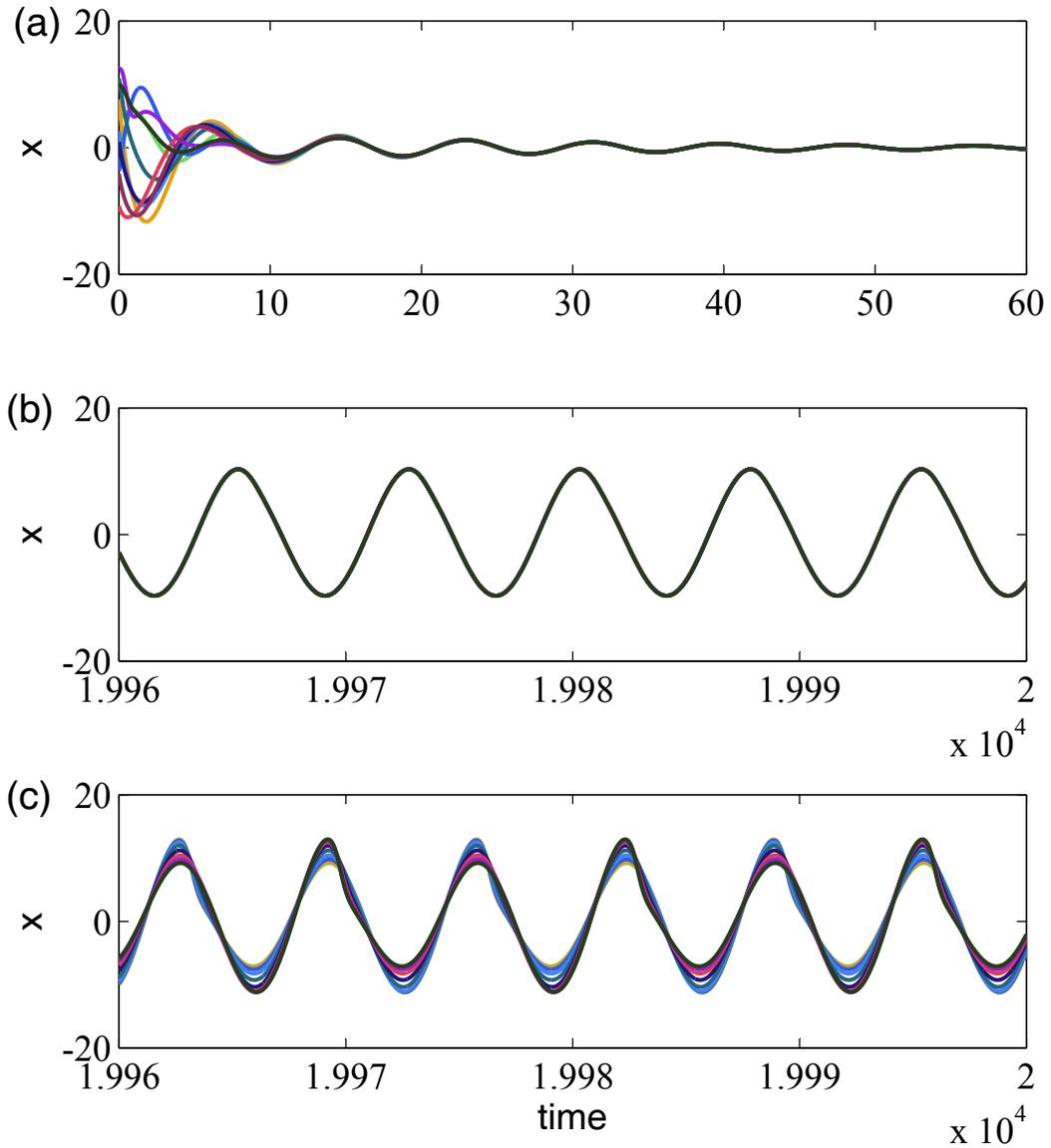} %
\caption{(Color online.) The evolution of $N=10$ oscillators indirectly coupled via heterogenous environment for various densities $\rho$ given $D=0.45$. (a) Oscillation death, $\rho=1.0$, (b) complete synchronous oscillation (periodic), $\rho=2.0$, and (c) phase synchronization (chaotic), $\rho=6.8$. }
\end{figure}

\begin{figure}[ptb]
\includegraphics[bb=1pt 322pt 401pt 776pt,clip,scale=1.0]{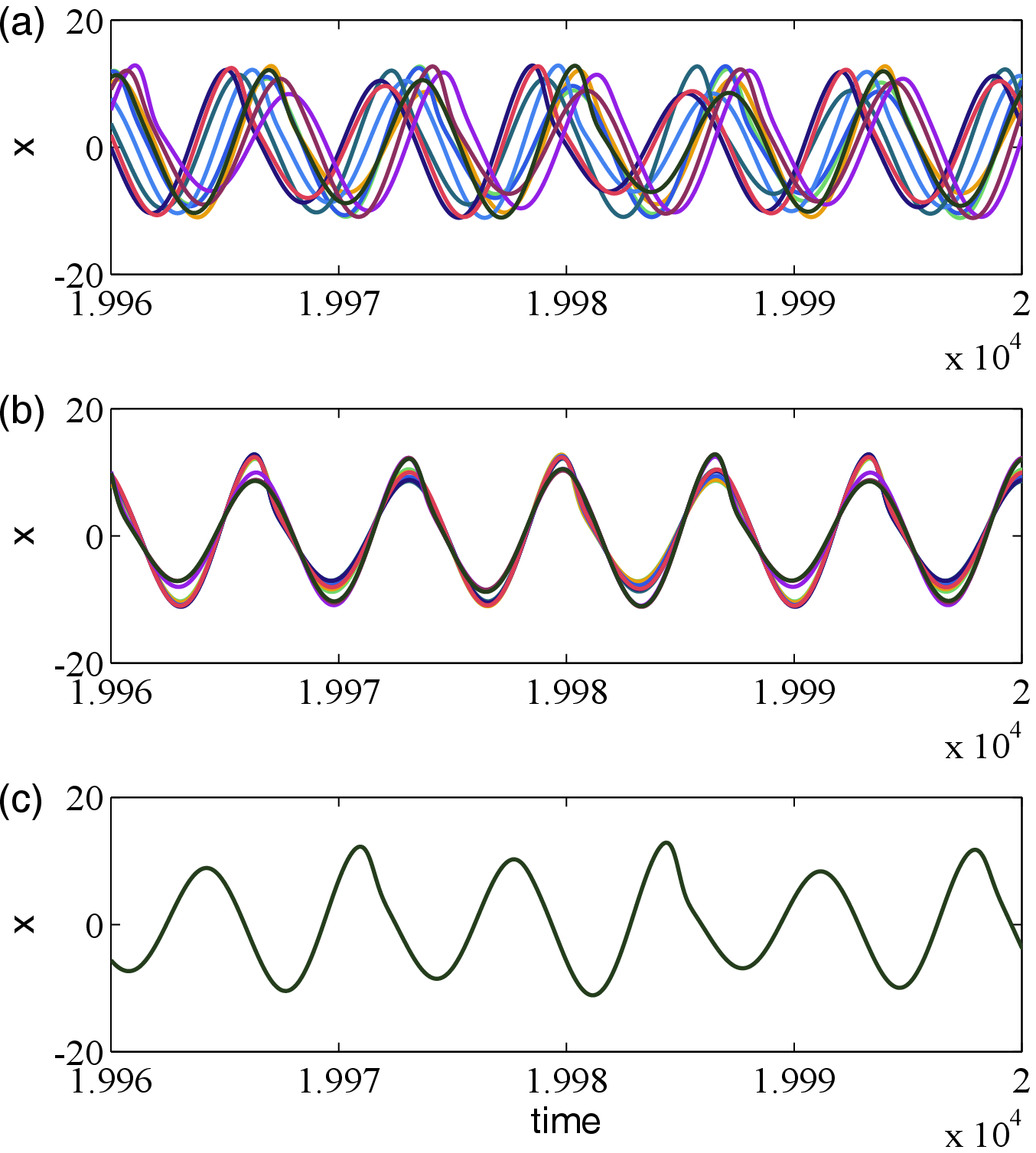} %
\caption{(color online) Similar to Fig. 2, but for various diffusion coefficients given $\rho=4.8$. (a) desynchronizaiton, $D=2\times 10^{-5}$, (b) phase synchronous oscillation, $D=0.03$, and (c) complete synchronous oscillation $D=0.30$. }
\end{figure}

Next, we will consider the synchronization behaviors of small populations of chaotic oscillators ($N=10$) that are arranged on a linear chain where each of them is indirectly coupled through a diffusive environment (refer to Fig.~1). Specifically, we are concerned the emergence of the various dynamical states (e.g., OD, PS and CSO) and how they switch from one state to another as we vary the population density and diffusion parameters.

Figure~2(a-c) shows three typical dynamical states of Eqs. (\ref{rossler}-\ref{env}) as a function of $\rho$ given $D=0.45$. With relatively small $\rho$ (e.g., $\rho=1.0$), the fully coupled system is damped to a stationary state where all of the oscillators keep quiescent after a short transient, see Fig.~2(a) where the traces of $x_{i},i=1,\cdots,N$ are shown. Such a state is also called the OD state. Note that in the absence of $K$, the subsystem Eq. (\ref{rossler}) shows self-sustain (chaotic) oscillation. It is thus that the occurrence of OD is intrinsically caused by the environment coupling. Increasing $\rho$ to beyond a critical value $\rho_{c}\approx 1.5$, CSO is eventually reached as shown in Fig.~2(b), and their oscillation can be either periodic (shown in the figure) or chaotic which depends on $\rho$. The transition from the stationary state to the collective oscillation is a kind of sudden behavior, resembling to QS transition, which has been found in experiments and theoretical works where a homogeneous dynamical environment is assumed \cite{monte,ojalvo,Taylorsci,toth}. Here, we show that QS is also possible even when the dynamical environment is heterogeneous.  An unexpected observation is that further increasing $\rho$ (using the same random initial conditions), will lead to degeneration of CSO to PS as illustrate in Fig.~2(c). To be concrete, the phase change coincidentally but the amplitude of oscillators seems to behave independently. Compared to CSO, PS is a weak form of the synchronization.

While, as we increase the diffusion parameter $D$, the system experiences a bit different process to synchronization, as illustrated in Fig.~3(a-c) where $\rho=4.80$ is taken. As it shows, when $D$ is sufficiently small, desynchronizing behavior is observed, see Fig.~3(a). For the small population of the system ($e.g., N=10$ considered in the present case), PS can be easily achieved with the order of $D\sim 10^{-4}$. Figure~3(b) shows a case of $D=3.0\times10^{-2}$ where PS happens. As we increase $D$ to beyond the critical value $D_{c}$, CSO could occur. For example, in Fig.~3(c) we show a CSO case for $D=0.3$. Unlike the behavior in Fig.~2(b-c), further increasing $D$ from $D_{c}$ will not lead to a transition from CSO to PS.
\begin{figure}[ptb]
\includegraphics[bb=-1pt 330pt 402pt 752pt,clip,scale=1.0]{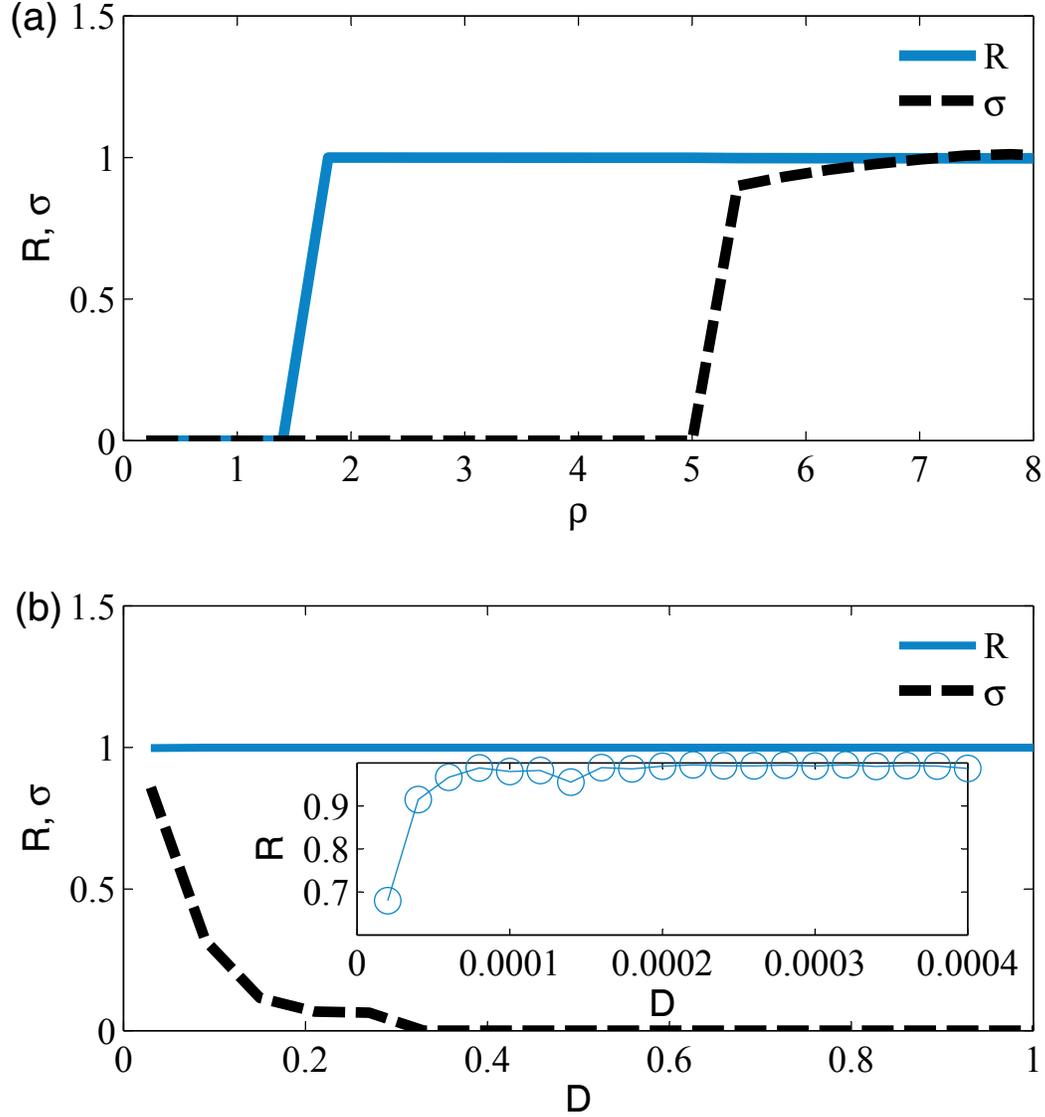}
 \caption{(color online) The synchronous indices $R$ and $\sigma$ as a function of (a) density $\rho$ and (b) diffusion coupling $D$, corresponding to Fig. 2 and Fig. 3, respectively. The inset plot in (b) shows the PS index $R$ as a function of $D$ in the regime of the small values.}
\end{figure}

How the synchronous states switch from one to another can be seen clearly by plotting the synchronization indices, i.e., $R$ and $\sigma$, as a function of the density and diffusion parameters. In Fig.~4(a), we show the change of $R$ and $\sigma$ as we increase the density $\rho$. For the PS index $R$, $R=0$ when the density is approximately less then $1.5$, which implies all oscillators are in quiescent state (or OD). With the increasing $\rho$, a sudden transition from $R=0$ to $R=1$ is observed which means the occurrence of synchronized oscillations. Such a transition is a typical QS similar to previous findings. In the OD state, CS index $\sigma=0$ as all the oscillators have the same values. (OD state can be seen a particular kind state of the complete synchronization state but here we distinguish it from CSO state.) Beyond the transition point, CSO can survive in a broad parameter of $\rho$ as indicated by $\sigma=0$. However, the CSO is degenerated to PS around $\rho \approx 5.20$ where $\sigma$ becomes a positive finite value and $R$ remains one.

We also plot $R$ and $\sigma$ as a function of $D$ in Fig.~4(b). The PS index $R$ keeps one for all used $D$ as the PS is reached with very small value of $D$ for this small system. This can be seen further from the inset where we plot the transition from desynchronous oscillation to synchronous oscillation as function of $D$. PS occurs in the order of $D\sim10^{-4}$. Unlike in the Fig.~4(a), the CS index $\sigma$ changes from non-vanished value to zero around the critical $D_{c}\approx 0.30$. It remains zero as we move $D$ onward.

To give a global picture of what roles played by the population density $\rho$ and diffusion constant $D$ in the synchronous behaviors of the indirectly coupled systems via a diffusive environment, we systematically calculated order parameter $R$ and $\sigma$ in the broad parameter regime of $\rho-D$ using different initial conditions. In Fig.~5 we show the distribution of $\sigma$ that is averaged over 81 random initial conditions as a function of $\rho$ and $D$. ( The distribution of the PS index $R$ (not shown here) behaves relatively simply, e.g., only two value $R=0$ and $R=1$ are detected and they are separated by the line at $\rho_{c}\approx 1.5$.) From Fig.~5, we can generally divide the phase-diagram into three regions: OD, CSO and PS. With fixed $K$, the OD region is only relies on $\rho$ and independent of $D$, but both regimes of CSO and PS depend on these two parameters. It is note that, given $D$, the system experience a process of OD $\rightarrow$ CSO $\rightarrow$ PS; while given $\rho$, the system experience a process of desynchronization $\rightarrow$ PS  $\rightarrow$ CSO. (Desynchronization regime is not shown in this phase diagram as it corresponds to the very small value of $D$.) Further more, the critical diffusion constant $D_{c}$ to achieve CSO is a function of $\rho$. The underlying mechanism of such transitions and dependence will be the focus of the following section.
\begin{figure}[ptb]
\includegraphics[bb=2pt 547pt 390pt 758pt,clip,scale=1.0]{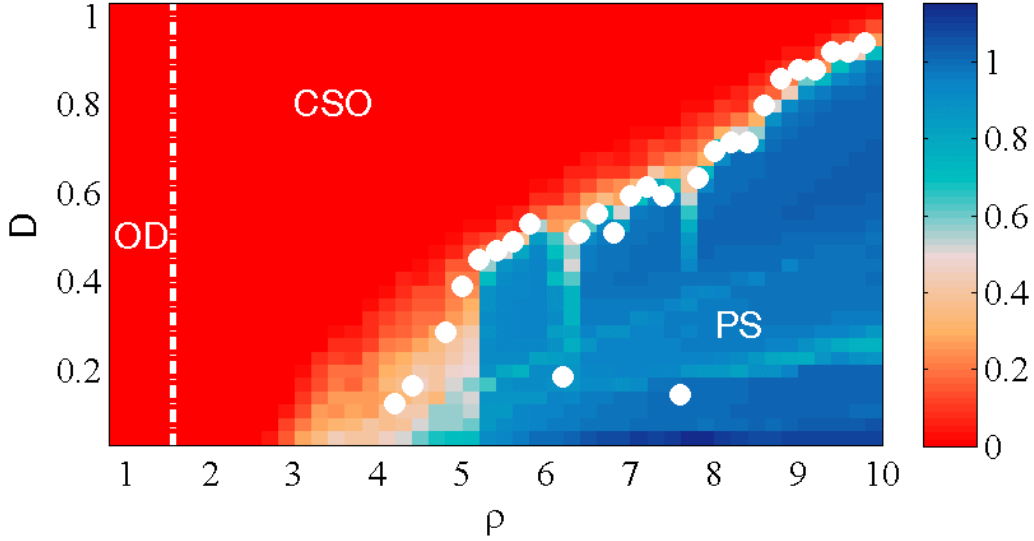} %
\caption{(color online). Phase diagram of dynamical states in the space of $\rho$ and $D$ indicated by the distribution of $\sigma$ which is averaged by 81 ensemble of initial conditions . The phase diagram is divided into three regimes: OD (oscillation death), CSO (complete synchronous oscillation) and PS (phase sysnchronization). The white dashed line is boundary between OD and CSO. The white dots are from MSF calculation.}
\end{figure}

\section{Mechanism analysis of the synchronization transition}

Before we step ward further, let's revisit the coupled system Eqs. (\ref{rossler}-\ref{env}). From the dynamical perspective, these equations represent an ensemble of oscillators indirectly coupled via a heterogeneous environment. The parameters $K$ denotes the coupling strength of oscillators described by $(x,y,z)$ coupled with dynamical environment $x^{e}$. The parameter $\rho$, representing the density of the population, also reflects the strength that the environment receives the information from the oscillators. Obviously, if either $K$ or $\rho$ is sufficiently small, synchronization is impossible. It is worth pointing out that there is an alternative explanation of these equations. For instance, we can treat the system as a new oscillator described by $(x,y,z,x^{e})$ and each new oscillator is coupled with the nearest ones. In this sense, $K$ and $\rho$ are the intrinsic parameters and $D$ represents the only coupling strength of this new oscillator system. In this sense, the $\rho$ plays not only strength but also intrinsic dynamics of the system, but $D$ only plays a coupling strength between the new oscillators described by $(x,y,z,x^{e})$. With this point of view, it is our expect that increasing $D$ would enhance the synchronization as we see in Fig.~3.

\subsection{The dual roles of $\rho$}
To better understanding the synchronous transition, we first investigate how the dynamics of the full system depends on $\rho$ when the spatial effects are excluded, i.e., $D=0$. For $K>0$, the oscillations behavior of the full system strongly relies on the population density $\rho$. For instance, for $K=1$, the fixed point $\mathbf{z}_{ss} = (x_{ss},y_{ss},z_{ss},x^{e}_{ss})$ is stable when $\rho$ is small and so there is no oscillatory behavior. By performing the linear stability analysis, we find that the fixed point solution will lose the stability around at $\rho_{c}\approx 1.50$ via a Hopf-bifurcation, which is consistent with the previous simulations. Crossing $\rho_{c}$, the system shows various self-sustain oscillations including periodic, quasi-periodic and even chaotic via a periodic-doubling bifurcation. Figure 6 shows a typical periodic-doubling bifurcation and corresponding the largest Lyapunov exponent as a function of $\rho$. These results strongly imply that the population density $\rho$ plays dual roles: increasing $\rho$ will (i) enhance the coupling between the oscillators and the dynamical environment, and (ii) lead a transition of the system from periodic to chaotic oscillations as well.

The synchronous transition can be viewed as the result of the competition of these two roles played by $\rho$. Qualitatively, when $\rho$ beyond but not far from the critical value $\rho_{c}$, the systems is periodic and in this case, CSO is observed. Increasing $\rho$ will lead the system to fall in the regime of the quasiperiodic or even chaotic, but it also increases the coupling strength. Therefore, CSO is still possible for a certain range of $\rho$. However, further increasing $\rho$, the systems falls to chaotic regime, and under this case $\rho$ as a role of coupling strength is not sufficient large to guarantee the emergence of the CSO. Consequently, we observe PS instead CSO for large $\rho$.

\begin{figure}[tbp]
\includegraphics[bb=1pt 312pt 432pt 528pt,clip,scale=1.0]{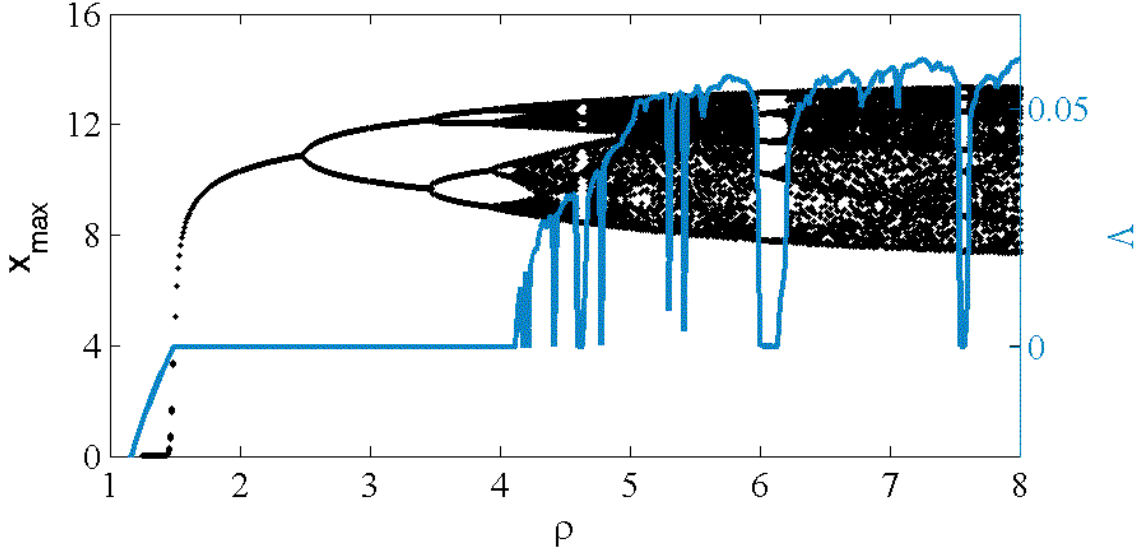} %
\caption{(color online). Periodic-doubling bifurcation and corresponding the largest Lyapunov exponent, denoted by $\Lambda$, of the full systems ($D=0$) as a function of the density parameter $\rho$ with $K=1.0$.}
\end{figure}

\begin{figure}
\includegraphics[bb=69pt 368pt 489pt 818pt,clip,scale=1.0]{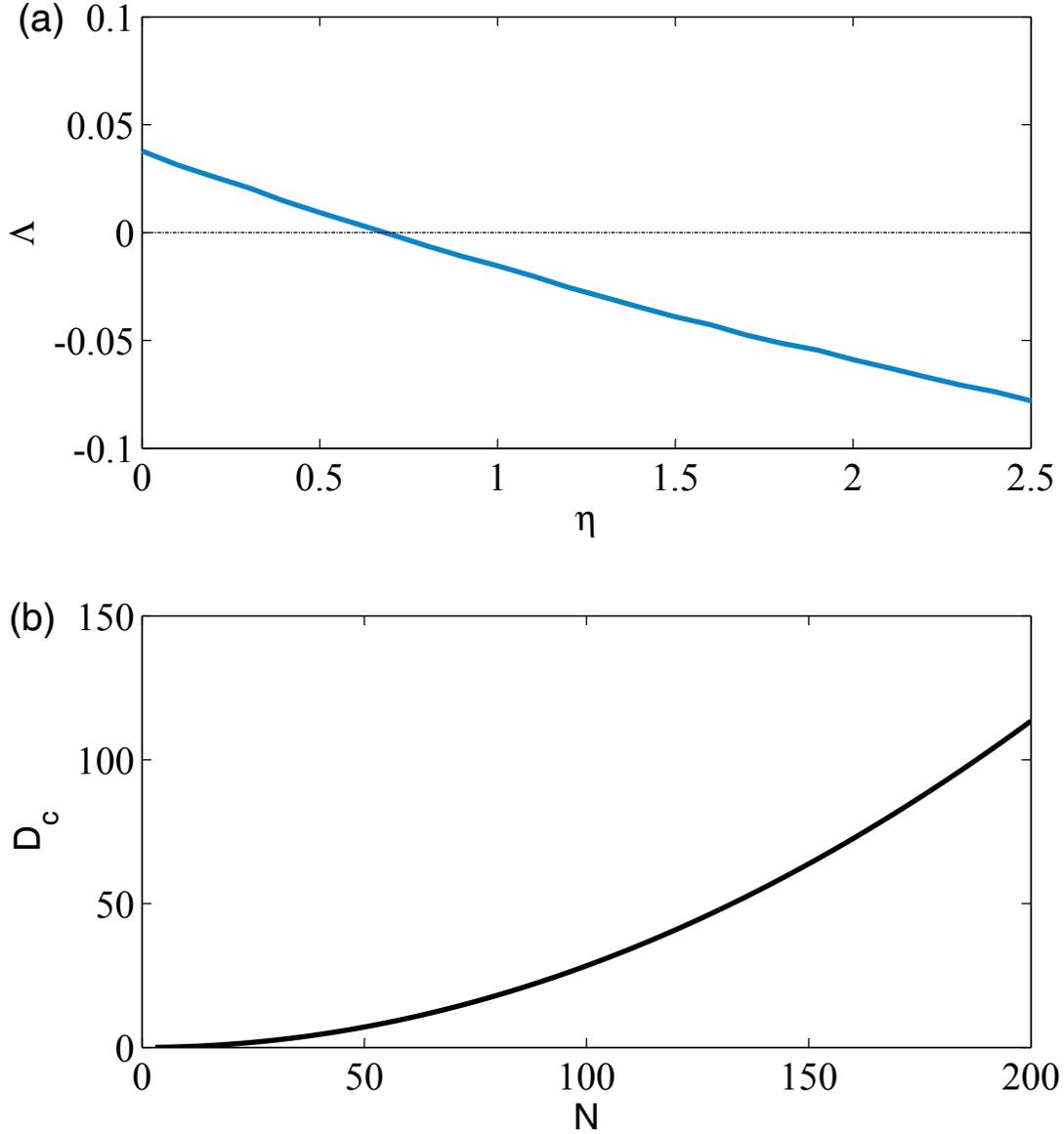} %
\caption{(color online). (a) The largest Lyaponov exponent $\Lambda$ as a function of normalized coupling strength $\eta$ for $\rho=4.8$. (b) The critical diffusion coupling $D_{c}$ as a function of the number of the oscillators $N$ with $\eta_{c}=0.7$, which corresponds to the critical value from (a). }
\end{figure}

\subsection{Master stability function approach}
To give deeper insights into the synchronous transition, particularly from CSO to PS as a function of $\rho$, we below using master stability function (MSF) \cite{percola,huang,fu} approach to quantitatively characterize the synchronous boundary separating these two regimes.

As discussed previously, Eqs. (3-4) can be viewed as a group of new oscillators coupled via the variable $x^{e}$ with the nearest coupling. $K$ and $\rho$ are the intrinsic parameters of the new oscillator system and $D$ represents the direct coupling strength. Also note that all the oscillators are assumed to be identical. In this sense, MSF is applicable to the present case.

Following the traditional MSF approach, we rewrite Eqs. (3-4) in a general form \cite{percola,huang,fu},
\begin{equation}
\frac{d\mathbf{x}_{i}}{dt}=\mathbf{F}(\mathbf{x}_{i})+\varepsilon\sum_{j=1}^{N}c_{ij}\mathbf{H}(\mathbf{x}_{j}),\label{msf1}
\end{equation}
where $\mathbf{x}$ denotes $d$-dimensional vector and $\mathbf{F}(\mathbf{x})$ is the local dynamics %
of the oscillator (the subscript $i$ is omitted here). $\varepsilon$ represents the coupling strength and %
$\mathbf{H}$ is a coupling function. $c_{ij}$ is the element of the coupling matrix $C$ which is %
determined only by the network topology, and satisfies $\sum_{j=1}^{N}c_{ij}=0$. We further assume %
that the coupling matrix $C$ has a set of real eigenvalues, say $\lambda_{i}$ ($i=0,\cdots,N-1$) and %
they are arranged in a following way, $0=\lambda_{0}>\lambda_{1}\geq\cdots\geq\lambda_{N-1}$.%

Let $\mathbf{x}^{s}$ be the synchronous manifold of the system (the manifold for the complete synchronization), i.e., $\mathbf{x}^{s}=\mathbf{x}_{1}=\mathbf{x}_{2}=\cdots=\mathbf{x}_{N}$. Now we consider an infinitesimal perturbation from the synchronous manifold, $\delta \mathbf{x}_{i}=\mathbf{x}_{i}-\mathbf{x}^{s}$. Substituting it to Eq. (\ref{msf1}) yields a variation equation
\begin{equation}
\frac{d \delta \mathbf{x}_{i}}{dt}=D\mathbf{F}(\mathbf{x}^{s})\cdot\delta\mathbf{x}_{i}+\varepsilon\sum_{j=1}^{N}c_{ij}D\mathbf{H}(\mathbf{x}^{s})\cdot\delta\mathbf{x}_{i},\label{msf2}
\end{equation}
where $D\mathbf{F}(\mathbf{x}^{s})$ and $D\mathbf{H}(\mathbf{x}^{s})$ are the $d \times d$ Jacobian matrices evaluated at the synchronous solution $\mathbf{x}^{s}$. Projecting $\{\delta x_{i}\}$ into the eigenspace spanned by the eigenvectors of the coupling matrix $C$, then the set of equations described by Eq. (\ref{msf2}) can be transformed to $N$ decoupled equations like \cite{fu},
\begin{equation} \delta\dot{\mathbf{y}}_{i}=\left[D\mathbf{F}(\mathbf{x}^{s})+\varepsilon\lambda_{i}D\mathbf{H}(\mathbf{x}^{s})\right]\delta\mathbf{y}_{i},\label{msf3}
\end{equation}
where $\delta\mathbf{y}_{i}$ represents the $i$th mode of the perturbations corresponding to the eigenvalue $\lambda_{i}$. The mode associated with $\lambda_{0}$ represents the motion parallel to the synchronous manifold, i.e., the trajectory of a single oscillator.

Denoting $\eta_{i}=-\varepsilon\lambda_{i}$($i=0,1,\cdots,N-1$) be a specific values of a normalized coupling parameter $\eta$, and the above equation can been seen as some particular cases of
\begin{equation}
\delta\dot{\mathbf{y}}=\left[D\mathbf{F}(\mathbf{x}^{s})-\eta D\mathbf{H}(\mathbf{x}^{s})\right]\delta\mathbf{y}.
\end{equation}
Let $\Lambda=\Lambda(\mathbf{x}^{s},\eta)$ be the largest Lyapunov exponent of the system. If $\Lambda$ is negative, an infinitesimal perturbation from the synchronous solution $\mathbf{x}^{s}$ will diminish exponentially and thus the solution $\mathbf{x}^{s}$ is stable, at least where oscillators initialized around the $\mathbf{x}^{s}$; while $\Lambda$ is positive, which means the small perturbation from the synchronous manifold will lead to the divergence of the trajectories, the synchronous solution $\mathbf{x}^{s}$ is unstable \cite{huang}.

For the system described by Eqs. (\ref{rossler}-\ref{env}), the coupling matrix is %
\[
C=\left(
  \begin{array}{ccccc}
    -1 & 1 & 0 & \cdots & 0 \\
    1 & -2 & 1 & \cdots & 0 \\
    \vdots & \vdots & \vdots & \vdots & \vdots \\
    0 & \cdots & 0 & 1 & -1 \\
  \end{array}
\right)
\]
and the corresponding Jacobian matrix is
\[
D\mathbf{F}=\left(
  \begin{array}{cccc}
    -K & -1 & -1 & K \\
    1 & a & 0 & 0 \\
    z^{s} & 0 & x^{s}-c & 0 \\
    K\rho & 0 & 0 & -K\rho-J \\
  \end{array}
\right)
\]
It has been shown that the matrix $C$ can be diagonalized and the eigenvalues are \cite{heagy},
\begin{equation}
\lambda_{k}=-4\sin^2\left(\frac{\pi k}{2N}\right),~ k=0,1,\cdots,N-1. \label{eigenvalue}
\end{equation}

In Fig.~7, we show the largest Lyapunov exponent $\Lambda$ as a function of the normalized coupling strength $\eta$ in the case of $\rho=4.8$. Typically, the MSF curve is monotonic and $\Lambda(0)>0$ as the full system show chaotic oscillation with $\rho=4.8$. Increasing $\eta$ will decrease $\Lambda$ and beyond some critical value $\eta_{c}\approx 0.70$, $\Lambda$ change its sign from positive to negative. To guarantee the emergence of the CSO, $-\varepsilon\lambda_{1}>\eta_{c}$, which requires critical diffusion parameter $D\geq D'_{c}=-\eta_{c}(\Delta x)^2/\lambda_{1}=0.286$. This value is close to $D_{c}=0.30$ obtained by the direct numerical simulation.

We systematically calculate the critical coupling strength $D_{c}$ for various $\rho$ at which the full system in the chaotic regime. The dependence of $D_{c}$ on $\rho$ is shown in Fig.~5 (see white dots). We find the results from the MSF approach agree quite well with the direct simulation of the full system.  From Fig.~5, we find that larger $\rho$ usually requires larger $D_{c}$ to achieve the CSO. Therefore, it is not a surprise finding that given the limit value of $D$, increasing $\rho$ could fail to CSO as we see in Figs.~2 and 3. Another implication from the MSF curve is that there is a limit size of oscillators to
achieve complete synchronization in the chaotic regime of Eqs. (\ref{rossler}-\ref{env}).
This is easily can be seen if we note that $D_{c}=-\eta_{c}(\Delta x)^2/\lambda_{1}$ with
$\lambda_{1}=-4\sin^{2}(\pi/2N)$. As $N \gg 1$, $D_{c}$ tends to be infinity to achieve CSO.
The dependence of $D_{c}$ on $N$ is shown in Fig. 7(b), and it clearly shows that increase
$N$ would dramatically increase $D_{c}$. Thus, the larger size of the system,
the more difficult to observe CSO, in the chaotic regime particularly.

\begin{figure}[tbp]
\includegraphics[bb=0pt 314pt 437pt 758pt,clip,scale=1.0]{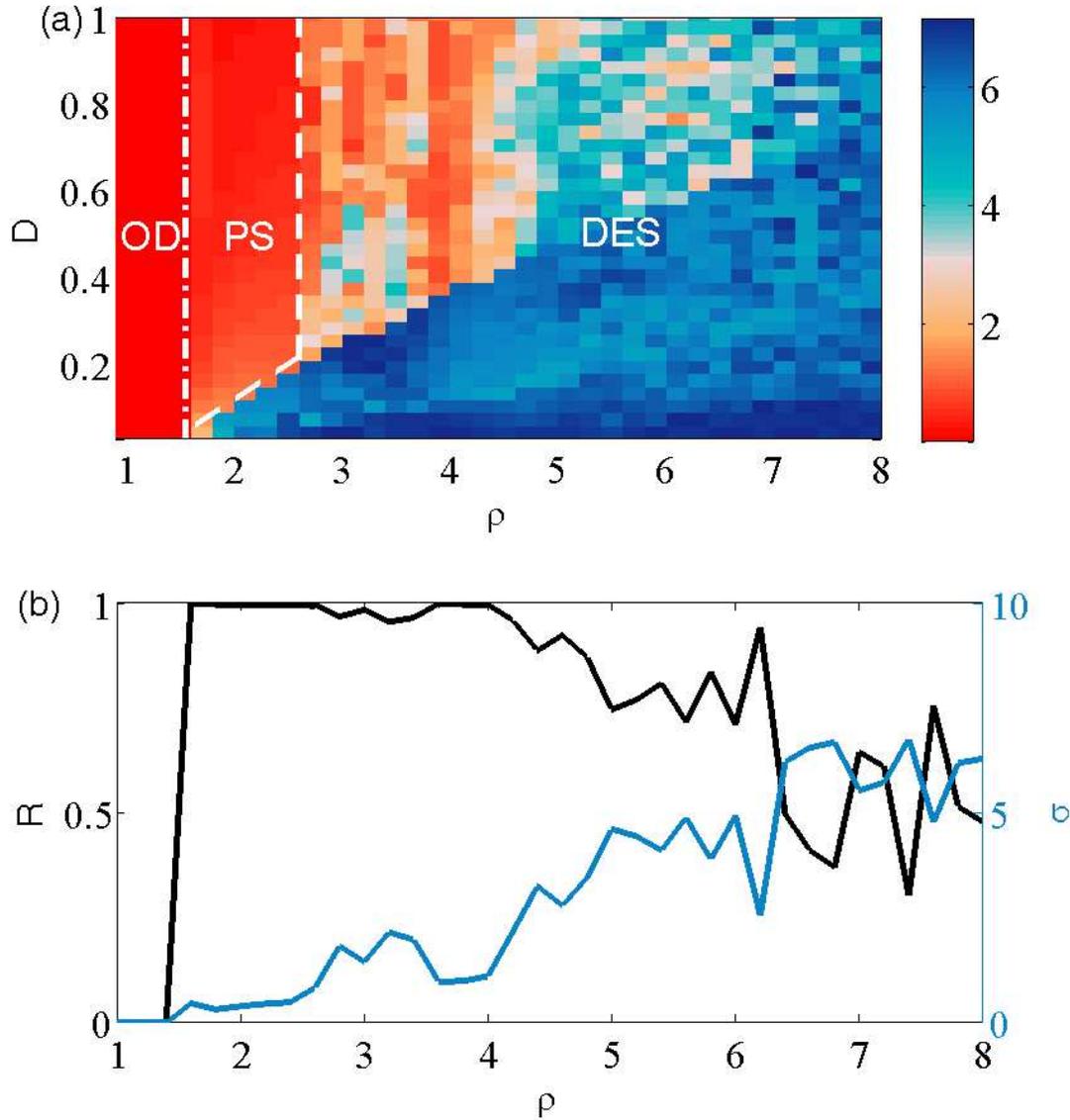} %
\caption{(a) Phase diagram indicated by the PS index for larger group of oscillators. OD: oscillation death, PS: phase synchronization and DES: desynchronization. (b) The synchronous indices $R$ and $\sigma$ as a function of density $\rho$ with $D=0.6$.}
\end{figure}

\section{Synchronization and traveling waves of large group of oscillators}
Till now, we have only considered the dynamical QS and its explanation for small size group of oscillators. As implicated by MSF approach, the size effects should be existed in our model. So in this section, we will consider the much larger size, e.g. $N=1000$, and see what kind of new behaviors could be observed.

An overview picture of the dynamical states indicated by the synchronization index $\sigma$ in the $\rho-D$ space is shown in Fig. 8 (a). With a comparison to Fig. 5, a significantly difference is that for large size system, the synchronization regime shrinks to a very narrow regime, and there is no complete synchronous oscillation regime and most of the regime is dominated by the desynchronous state. An interesting finding is that increasing $\rho$ first leads to the PS, but continuous to increase $\rho$ would then lead to the desynchronization, which is not observed for indirectly coupled oscillators via a homogeneous environment. The above non-traditional quorum sensing transition is clearly seen in Fig. 8(b).

\begin{figure}[tbp]
\includegraphics[bb=59pt 547pt 414pt 728pt,clip,scale=1.2]{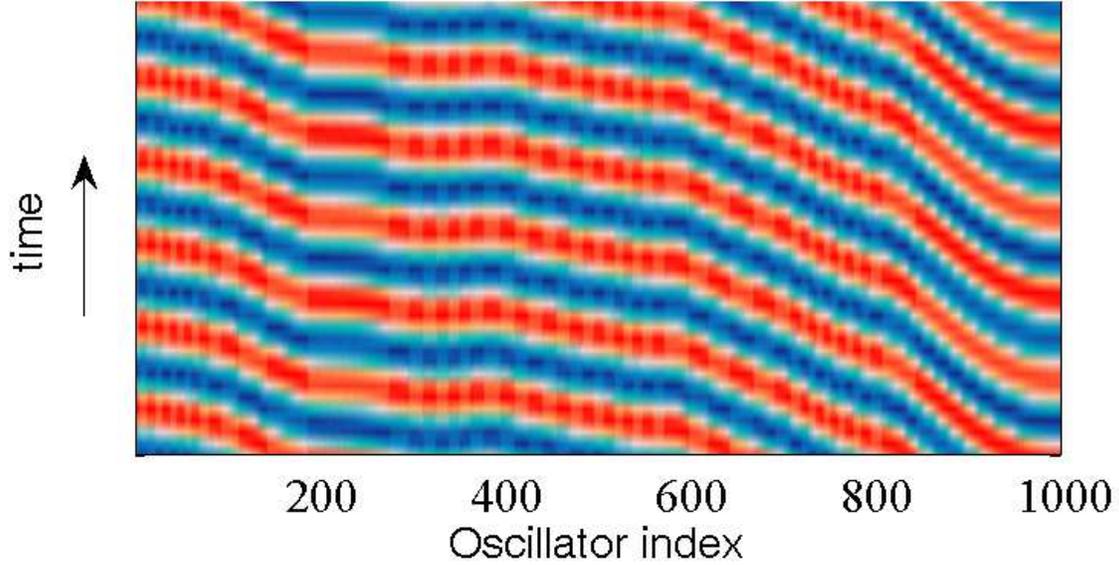} %
\caption{(color online) Traveling wave emerges in the DES regime in the case of large population oscillators. $D=0.03$ and $\rho=3.0$.}
\end{figure}
Finally, we would like to point out that even in the desynchronization regime, the oscillators do not oscillates completely disorder. In contrast, they sometimes may show somehow coherent structure. For instance, traveling waves could also emerge. In Fig. 9, we show a spatiotemporal plot of the evolution of the one thousand oscillators, a traveling wave propagates from the
right to left. For larger $\rho$, the system become less coherent and partial traveling waves
would occur in the system (no figure shown). The wave patterns have been reported in previous
work such as the glycolyses \cite{schutze_bj11} and genetic network \cite{danino}, however, our finding shows the first
evidence of the irregular waves observed in populations of chaotic oscillators coupled
via a heterogeneous environment.

\section{Discussion and Conclusion}
By coupling R\"{o}ssler chaotic oscillators to a heterogeneous environment, we have performed a systematic investigation on collective behaviors in such a indirectly coupled system. The present work is a natural and nontrivial extension to the indirectly coupled oscillators with the homogeneous dynamical environment. Although the QS is quite common in oscillator systems with homogeneous environment, but whether it is extended to a population of spatially distributed chaotic oscillators indirectly coupled through a heterogeneous environment remains poorly known. Previous work on QS give us an impression that larger the population density, the easier the system become synchronization. In this work, we uncover a different scenario: the degree of synchronization decrease after the onset of QS transition as we further increase the population density. Specifically, a degeneration from CSO to PS (for small size, e.g., $N=10$) or from PS to desynchronization (for large size, e.g., $N=1000$) could occur as we increase the population density. This scenario has not been reported in previous work. On the other hand, the present work is also an essential complement to QS study in a population of oscillators coupled via a heterogeneous environment. For instance, in pervious study of heterogeneous environment, they usually were concentrated on the formation of the spatiotemporal patterns \cite{danino,dilanji_jacs12,schutze_bj11,noorbakhsh_pre15}, and the fundamental problems such as the onset of synchronization and its dependence of the density is still largely understudied. What's more, they only consider the periodic oscillator rather than the chaotic case \cite{dilanji_jacs12,schutze_bj11,noorbakhsh_pre15}, the latter is closer to the natural systems.

It is noted that we only presented our results with the no-flux boundary for the dynamics of the extracellular solution of the system in the text. Additional studies were also performed on the ring of the chaotic oscillators where the periodic boundary conditions were applied and similar results were observed. The only difference is that the critical value of the diffusion coupling $D_{c}$ is bit smaller than the no-flux boundary case. This could be explained by noting that the eigenvalue of the periodic case is much smaller that of the no-flux boundary for the same system parameters.

Our findings may be tested by various chemical or biological experiments. For instance, in biological systems such as yeast cells, it has been shown that under certain conditions the glycolytic oscillation of the yeast could be chaotic \cite{nielsen}. This makes it possible to design a similar experiment likes one used in Ref. \cite{schutze_bj11}. Another candidates may be the chemical realization of the system. For instance, it is already to generate periodic Belousov-Zhabotinsky (BZ) droplets \cite{toiya} or particles \cite{Taylorsci} which can be fixed in space. Also, it is known that chaotic behaviors can be observed in BZ chemical reactions \cite{schmitz,rossi,mindlin}. In this sense, it may be possible to make a similar configuration like Fig. 1 in experiment with chaotic BZ particles. For these possible experiments, it will be interesting to test the synchronization transition as well as check the bifurcation of the dynamical systems as it coupled to the heterogeneous environment.

In summary, we have made a systematic study of QS behaviors in a population of chaotic R\"{o}ssler oscillators indirectly coupled through a diffusive dynamical environment. We observed various dynamical states involving OD, PS as well as CSO in the parameter space expanded by $\rho-D$. A non-traditional QS transition was also observed and such unexpected QS was found due to the competition of the dual roles played by the population density. The separation boundary between the CSO and PS could be numerically predicted by the MSF approach by simply treating the full systems as a locally coupled oscillator system. The travelling waves were also possible in the large size of the system.  We finally made a brief discussion on the possible realization in experiment on our proposed system.\begin{center}
\textbf{ACKNOWLEDGMENT}
\end{center}
This work was supported by the National Natural Science Foundation of China under Grant Nos.
11205039 and 11505153, Natural Science Foundation of Zhejiang Province under Grant Nos. LY16A050003 and LQ15A050002, and the funds from Hangzhou City for the Hangzhou-City Quantum Information and Quantum Optics Innovation Research Team.

\end{document}